\documentstyle[12pt,aasms4]{article}

\begin{document}
\def \m {$ M_\odot$~}
\def \l {$ L_\odot$}
\def \ro {g cm$^{-3}$ }
\def \etal {\it et al. \rm}

\title{ Delayed Detonation at a Single Point in Exploding White Dwarfs}
  
\author{ E. Livne }
\affil{Racah Institute of Physics, The Hebrew University, Jerusalem 91904, Israel}
\authoremail{eli@frodo.fiz.huji.ac.il}

\begin{abstract}

Delayed detonation in an exploding white dwarf, which propagates from an off-center
transition point, rather than from a spherical transition shell,
is described and simulated. The differences
between the results of 2D simulations and the 1D case are presented and discussed.
The two dimensional effects become significant in transition density below $3 \times \rho_7$,
where the energetics, the production of Fe group elements and the symmetry of the explosion
are all affected. In the 2D case the explosion is less energetic and less Ni is produced in
the detonation phase of the explosion. For low transition density the reduction in Ni mass can
reach 20-30 percent. The asymmetry in abundances between regions close
to the transition point and regions far from that point is large, and could be a source
to polarization patterns in the emitted light. We conclude that the spatial and
temporal distribution of transition locations, is an important parameter which must be included
in delayed detonation models for Type Ia supernovae.  
\end{abstract}

\keywords{hydrodynamics-nucleosynthesis,abundances-supernovae:general} 

\section {Introduction}

  The delayed detonation (DD) model is a successful model for type Ia supernovae (SnIa),
 as it meets most of the constraints imposed by observations (\cite{kok91a}, \cite{kok91b}). 
The DD model assumes a transition from a deflagration to a detonation
(DDT) at some critical density (of the fuel ahead of the front), after a 
significant expansion of the WD during a slow deflagration phase. This critical
density is limited to a fairly small range around $2\times 10^7$ \ro by
observational constraints (\cite{hof95a},\cite{hof95b},\cite{hof96}). It is also
the density at which turbulence can destroy the flame (\cite{KOW97}). In this paper
I do not discuss the difficult problem of the DDT itself, but
rather examine the geometrical effects of asymmetrical transition. 
The vast majority of numerical simulations that have been performed
up to now assumed  spherical symmetry, where the transition
to detonation occurs in a spherical shell. Only a few simulations have tried
to include multi dimensional effects (\cite{arliv94a}, \cite{arliv94b}), but
in those works the geometrical effects of the location of the transition, on the
explosion, were not considered. One can judge different model by comparing the
masses and velocities of different products to observations. In particular,
the chemical composition of the ejecta versus velocity is an important imprint of any
model. A typical common problem to  spherical DD simulations is a small,
 but not insignificant, amount of slow
moving intermediate mass elements (3000-4000 km/sec), around the transition
point (\cite{hof95a}, \cite{til99}). Such slow moving intermediate mass elements are
hard to observed, and they do not show in theoretical spectra.

The purpose of this letter is to go beyond the 1D case by studying the more probable
scenario, in which the transition to detonation occurs off-center
in an isolated point, located near on the deflagration surface. Then the
 hydrodynamical problem is not spherically
symmetric, but rather axially symmetric around the line which passes
through the center and the transition points. In exploring this scenario we are motivated
by two facts - first, when the density ahead of the front drops to the critical density the
radius of the (spherical) front is roughly a third of the total radius, and second,
DDT occurs on scales much smaller than the radius of the star, or
even the radius of the front (\cite{KOW97}). Hence, transition at a point is a good approximation
to any realistic scenario. Moreover, there are still severe doubts as to
the robustness of DDT in thermonuclear flames (\cite{nim99}), so DDT in a spherical shell
of such a large radius seems very improbable. The case we are studying here, off-center point
transition, is the
most extreme asymmetric case, while transition in a spherical shell is the most symmetric
case. Assuming one point transition,
The detonation front propagates from the transition point inside the unburnt shell
and around the already burnt core, in a manner similar to the detonation in a helium
shell (\cite{livgla90}). Although the time needed for this detonation wave
to consume the whole star is much shorter than any other dynamical time (roughly
 0.5sec), we can not rule out the possibility of DDT in several unsynchronized points.
This however would require 3D simulations which are currently beyond our
computing capabilities. Regarding the symmetry of the explosion, transition in several
points would behave as an intermediate case between the spherical case and the single 
off-center case. I shall refer to this possibility in the concluding remarks
at the end of the paper. In the next two section we describe briefly the
hydrodynamics, the simulations and mainly the observational results of this
 unexplored scenario.

\section {The models and the numerical techniques}

The simulations are performed using the 1D and 2D versions of the hydro
code VULCAN (\cite{liv93}), supported by an accurate and tabulated equation of
state and a reaction network that is based upon the \it alpha network \rm (Thielemann, private
communication). The initial model consists of a cold, almost isentropic WD
of 1.39 \m, central density of 2$\times \rho_9$ and a constant C-O ratio of 1. The deflagration
 phase is performed in a 1D spherically symmetric
simulation, using a front speed of 2 percent of the local sound speed (\cite{kok91a}).
In coarse grids there are nearly 200 uneven radial zones, where the resolution around the transition
radius is $\delta m = 0.01 \times$ \m, which corresponds to a zone size of 20 km. 
We consider 3 cases of transition densities - $\rho_{tr}=2\rho_7$ (case A), $\rho_{tr}=3.5\rho_7 $
 (case B) and $\rho_{tr}=5\rho_7$ (case C), for which the deflagration front radius is 
$r_f=$1880km, 1500km and 1300km respectively.
For each case, the deflagration stage terminates when the density ahead of the front reaches
the critical density. 

In order to be able to compare the two cases I deliberately choose to start the detonation phase with
a spherical undisturbed configuration, that is produced by the 1D code. By simulating the deflagration
phase in 1D we neglect the complicated structure of the front due to Rayleigh-Taylor instability.
Hence, the results presented in this work must not be considered as a new realistic DD model. They
measure only the effects of the geometry of the transition on the explosion products. This simplifications
will be removed in future work, where the deflagration phase will be done in 3D. The location of the
transition point, however, will remain a free parameter, as the physics of the transition is not well
understood, and in any case DDT occurs on scales much smaller than the resolution of any realistic 
simulation.

At the end of the deflagration stage the density behind the front is almost
half $\rho_{tr}$. In the  lower density cases, intermediate mass elements (Si) burn very
slowly and the typical Si-scale of the front exceeds 100 km (\cite{gamezo99}).
In 1D simulations the transition to detonation at the front pushes the unburnt matter
ahead of the front forwards, and emits a rarefaction wave into the partially
burnt matter behind it. As a result, burning of intermediate mass elements
near the transition radius is never completed. In our 1D simulation
the large scales (Si scales) of the front ($\approx 100 km$) around the transition
radius are not finely resolved, but the results with finer zones are not very different.
The transition to detonation is done numerically by increasing the front speed
around the transition point (the intersection of the front with the right axis in fig 1a)
 to the sound speed.  I simulate the detonation phase
both in 1D, in 2D with shell ignition (a validity check) and in 2D with DDT at a single
point. In the 2D grid it takes about 20-30 zones for
the sliding front to reach a quasi stable Chapman-Jouguet (CJ) state. The corresponding
distance from transition to a CJ state is much larger in the simulation than in realty,
where it is known from experiments that this distance is roughly 10,000 times the
width of the front. The 2D simulations are terminated 1 second after DDT, where all burning
processed are already completed. Relatively small changes in abundances distribution are
possible after this time.

\section {Results from 2D simulations}

The main structure of the detonation front in case A is seen in figure 1a. The most obvious
feature, which is absent in 1D simulations, is  the inward moving oblique shock, induced
by the sliding detonation. This shock raises the density of the partially burnt
fuel, below the front radius, by almost a factor of 2. However, despite a significant
enhancement of the burning rates, intermediate mass elements are not exhausted.
When the detonation front converges to the axis of symmetry on the far side it becomes
stronger, and a Mach reflection is formed after it meets the axis. This causes the ejection
of a jet of burnt matter, mostly Ni, along the axis.
In all three cases, less iron group, and more silicon group species, are produced in the 2D simulation,
compared to the 1D case. The Ni productions in the 1D simulations for cases A,B,C are 0.68\m, 0.97\m and
1.07\m, while in the 2D simulations the corresponding numbers are 0.44\m,0.85\m and 0.99\m. Si productions
in the 1D cases are 0.27\m,0.16\m and 0.12\m while in the 2D cases they are 0.36\m,0.19\m and 0.14\m.
We see that the two dimensional effect becomes stronger with decreasing transition density.

\placefigure{fig1}

There are two interesting hydrodynamical effects which work
together towards a less energetic explosion and less efficient combustion. The first
one is related to the direction of the detonation front with respect to the density
gradient. In the spherical case, the detonation is not a CJ detonation but a \it strong
 detonation \rm. This is a result of the decreasing density gradient which accelerates
the shock beyond its CJ values. In the 2D case, the angle between the detonation front
and the density gradient varies from point to point. Close to the nearly stalled deflagration
front the detonation slides along the burnt core in a constant density background, so
it travels with almost CJ states. Even far from the transition shell the detonation is
not oriented in the radial direction (see figure 1a), so it is weaker than the spherical
detonation front.

 The second hydrodynamical effect is the continuous expansion of the
star, during the travel of the detonation front around the transition front. It turns
out that the expansion velocity near the front (nearly 800-1000 km/sec) is enough to
cause an increase of the front radius by 20-25 percent during the travel of the detonation
 around the burnt core. Since in a subsonic expansion the density changes like 
$r^{-3}$, the fuel density head of the detonation front drops by almost a factor of two
during this time, and consequently the burning temperature is also lower. As a whole, the fuel
density ahead of the detonation drops continuously
as it travels around, and by the time it reaches the far side the fuel density is roughly
half its original value. In addition, this effect causes
a pronounced asymmetry in the abundances of Fe group elements between the near and far sides
along the axis. The effect is clearly seen in figure 1b, in which we present
the average atomic number at the end of the 2D simulation of case A.

\section {Conclusions and Discussion}

The scenario of deflagration to detonation transition (DDT) in a
single off-center point, rather than in a shell, have been considered. In order to be able to compare
the two cases I deliberately choose to neglect the Rayleigh-Taylor instability during the deflagration
phase, and to start the detonation phase with a spherical undisturbed
configuration, that was produced by the 1D code. Although this way imposes unrealistic initial
conditions for the detonation phase, it enables one to isolate the net effect of the geometry of
the transition. 
Hence, the results presented in this work must not be considered as a new realistic DD model. They
measure only how the geometry of the transition affects the explosion products.
The results show  significant differences between the 1D case and the 2D case,
in the amounts and in the distributions of the explosion products. This implies that
the spatial and temporal distribution of DDT points is an important parameter for
modeling SnIa's. In the case of DDT in
an off-center point, located on the deflagration front, less Ni is produces compared
to the spherical case, as a result of a weaker detonation front and the continuously
decreasing density background. The difference increases as the transition density
becomes lower, and can reach a reduction of 20-30 percent in the Ni production at $\rho_{tr}=2\rho_7$.
 In addition, the 2D effects tend to widen the range of intermediate
mass elements in velocity space, and would lower the minimal velocity for such elements.
This is an important consideration for a consistent comparison between models and
observations (\cite{til99}). For low transition density ($2 \times 10^7$ \ro)
the intermediate mass elements produced at the end of the deflagration phase do not
disappear in the 2D simulation. The oblique shock induced by the sliding detonation upon
the burnt core is just too weak. This conclusion however needs verification with finer
zones than used here, near the edge of the deflagration front. 

Another important outcome of the scenario is a pronounced asymmetry in the Ni
abundance along the axis, and again, this asymmetry increases with decreasing transition
density. Recent analysis of polarization in some  SnIa's provides evidences for asymmetry
in the explosion products (\cite{wang97}). Modeling those polarization patterns suggest that
the asphericity is attributed to a region between a slow moving, almost spherical, Ni core
and a spherical Si layer. This is exactly the feature seen in fig 1b, where the products of the
deflagration phase form a spherical Ni core, and the products of the detonation phase
form aspherical region of partial burning, above which lies an almost spherical Si shell.
For transition density above $3.5 \times 10^7$ \ro  the asphericity almost disappear except
for the narrow jet on the far axis. Consequently, the scenario of DDT in a single point can be founded
if a relation between Ni mass and asphericity (polarization) will be found in observations.
Theoretically, a more systematic study of the range $ 2\rho_7 < \rho_{tr} < 3\rho_7 $ is required.
 
It is clear that in the scenario presented here, the transition density that would
give a best fit to observations is higher than in the 1D case. This however will be
different in the pulsational delayed detonation model (PDD), if DDT occurs during the
contraction phase of the pulse. Thus, we expect that the differences between the
direct DD and the PDD models will be larger in the scenario of point transition than
in the case of a spherical transition.
Several other complications may alter somewhat the results. First, in reality and in more
realistic multi dimensional simulations of the deflagration phase, the deflagration front will be
distorted and eventually develop disconnected burning pockets (\cite{nim97}). This will cause
 the detonation front to move in an inhomogeneous medium and complicated features of shock
interactions are possible, mainly close to the edge of the deflagration front.
The main uncertainty still is the location of the transition point, and even in future 3D
simulations this location (s) will remain a free parameter that strongly affects the explosion products.  

Secondly, several points may run to detonation before the first detonation wave has completed
its propagation around the star. This effect can alter somewhat the lowest modes
in the angular expansion of the asymmetry. It will require synchronization of the transition
at different points to a time interval of roughly 0.5 second. The interaction between detonation waves
that propagate from different transition points would form regions of enhanced burning
(Mach branches), and the effect is expected to produce much larger inhomogeneity than
the instability described in \cite{gamezo99}. The effect can disappear
only if the distance between different and simultaneous transition points, is less than $\L_{crit}$ -
the characteristic distance needed to form a stable CJ detonation. In our case, $L_{crit}$
is of the order of several meters up to a few kilometers (\cite{KOW97}), so one needs thousands
of simultaneous transitions in order to suppress shock interactions. This seems very improbable.
Finally, both complications mentioned above, and many other factors of the DDT mechanism,
act in a chaotic manner and will give rise to significant variations between SnIa events.

\acknowledgments

This work was motivated by P.H\"oflich during my visit to Aspen Center of Physics
in summer 1999. I also thank P.H\"oflich for enlightening comments and suggestions
to the first draft.

\clearpage

\figcaption[fig1.eps]{ a-(up)   Pressure during the detonation phase,
                       b-(down) Average atomic number at t=1 sec after DDT \label{fig1}}

\end{document}